\documentclass{ws-ijmpd}

\begin{document}

\markboth{M. Ujevic, P. S. Letelier and D. Vogt} {Relativistic Ring 
Models}

\title{RELATIVISTIC RING MODELS}

\author{MAXIMILIANO UJEVIC}

\address{Centro de Ci\^encias Naturais e Humanas, Universidade Federal 
do ABC, 09210-170 \\ Santo Andr\'e, S\~ao Paulo, Brasil \\ 
mujevic@ufabc.edu.br}

\author{PATRICIO S. LETELIER}

\address{Departamento de Matem\'atica Aplicada, Instituto de 
Matem\'atica, Estat\'{\i}stica e Computa\c{c}\~ao Cient\'{\i}fica, 
Universidade Estadual de Campinas, 13081-970 \\ Campinas, S\~ao Paulo, 
Brasil \\ letelier@ime.unicamp.br}

\author{DANIEL VOGT}

\address{Departamento de Matem\'atica Aplicada, Instituto de 
Matem\'atica, Estat\'{\i}stica e Computa\c{c}\~ao Cient\'{\i}fica, 
Universidade Estadual de Campinas, 13081-970 \\ Campinas, S\~ao Paulo, 
Brasil \\ dvogt@ime.unicamp.br}

\maketitle

\begin{abstract}

\hspace{1.5cm} Dedicated to the memory of Professor Patricio S. Letelier
\vspace{0.2cm}

\noindent Relativistic thick ring models are constructed using 
previously found analytical Newtonian potential-density pairs for flat 
rings and toroidal structures obtained from Kuzmin-Toomre family of 
discs. In particular, we present systems with one ring, two rings and a 
disc with a ring. Also, the circular velocity of a test particle and its 
stability when performing circular orbits are presented in all these 
models. In general, we find that regions of non-stability appear between 
the rings when they become thinner.
\end{abstract}


\section{Introduction} 

Ring structures are common features in different astrophysical objects 
on different scales. In particular, we have ring structures in the giant 
planets of the Solar system and also in several ring galaxies (or R 
galaxies) which are objects with approximate elliptical rings and no 
luminous matter visible in their interiors.\cite{the:spi,the:spi2} 
Furthermore, sometimes, as the result of interactions between galaxies, 
a ring of gas and stars is formed and rotates over the poles of a 
galaxy, resulting in the polar-ring galaxies.\cite{whi} In Newtonian 
gravity, the potential of a ring enclosing a disc can be obtained by a 
process of complexification of the potential of a punctual 
mass.\cite{app,whi:wat,gle:pul} This potential was used to build a 
family of similar structures.\cite{let:oli} Also, the Lord Kelvin 
inversion theorem\cite{kel} can be used to invert a member of the Morgan 
\& Morgan family of discs\cite{mor:mor} to produce an infinite disc with 
a central hole of the same radius as the original disc. We can define a 
cut-off in these inverted disc to represent flat rings. Furthermore, 
Letelier constructed several families of flat rings using the 
superposition of Morgan and Morgan discs of different 
densities.\cite{let} Recently, Vogt \& Letelier\cite{vog:let} found 
potential-density pairs representing flat ring structures by superposing 
members of the classical Kuzmin-Toomre family of discs.\cite{kuz,too} By 
using a suitable transformation, they found that these flat ring systems 
can be used to generate three dimensional potential-density pairs with 
toroidal mass distribution. The main advantage of these toroidal mass 
distribution models is that the density and the gravitational potential 
are given in terms of elementary functions. Also, similar toroidal 
analytical systems have been obtained using a different technique, known 
as the complex-shift method.\cite{cio:gia,cio:mar} In general 
relativity, the properties of space-times that represent superpositions 
of a Schwarzschild black hole and a disc or ring were investigated by 
several authors.\cite{cha,lem:let,sem:zac,sem:zel} The purpose of this 
article is to consider three-dimensional models for the gravitational 
field of rings and disc with rings in the context of general relativity 
following the method used in Ref. \refcite{vog:let2}. As a very first 
test for stability, we study the stability of particles in circular 
orbits using the relativistic version\cite{let2} of the Rayleigh 
criterion of stability.\cite{ray,lan:lif} We know that a better approach 
to study its stability is by performing first order perturbations of the 
energy-momentum tensor that take into account the collective behavior 
of the particles.\cite{uje:let,uje:let2,uje:let3} This study will be 
done in a future work.

The article is organized as follows. In Section 2, we present a special 
form of the isotropic Schwarzschild metric and the components of the 
energy-momentum tensor as functions of the metric coefficients, from 
which the physical properties of the matter distribution can be 
calculated. We also write expressions for the tangential circular 
velocity and specific angular momentum of test particles in circular 
motion. In Section 3, we apply the results of Section 2 to a single ring 
system, a disc with a ring system and a double ring system, which are 
particular cases of a family of solution found in Ref. 
\refcite{vog:let}. Finally, in Section 4, we summarize our results.

\section{Einstein Equations}

In General Relativity are two methods to find exact solutions of 
Einsten's field equations. The first method is the so called direct 
method, in which we set the energy-momentum tensor and then we find the 
metric tensor. In the other hand, most of the models (for example, disks 
and rings in astrophysics) were found by using the inverse method, in 
which we use the metric to calculate its energy-momentum tensor. In this 
article we used the second. But, not every solution found using this 
method is physical acceptable. The models must satisfy the energy 
criteria and also is desirable to have a clear physical interpretation 
at the newtonian limit. So, the justification of the method is done a 
posteriori and it will depend of the outcome. With this in 
consideration, let us consider the particular case of axially symmetric 
spacetimes in isotropic coordinates
\begin{eqnarray}
ds^2 = \left( \frac{1-f}{1+f} \right)^2 c^2 dt^2 - (1+f)^4(dR^2 + dz^2 + 
R^2 d\varphi^2), \label{metric}
\end{eqnarray}

\noindent where $R$, $\varphi$ and $z$ are the usual cylindrical 
coordinates and $f=f(R,z)$. From Einstein equations, $G_{\mu\nu} = -(8 
\pi G/c^4) T_{\mu\nu}$, and the metric (\ref{metric}) we obtain the 
following expressions for the components of the energy-momentum tensor
\begin{eqnarray}
T^t_t &=& - \frac{c^4}{2 \pi G (1+f)^5} \left(f_{,RR} + f_{,zz} + 
\frac{f_{,R}}{R} \right), \label{TTT} \\
T^R_R &=& \frac{c^4}{4 \pi G (1+f)^5 (1-f)} \left(f f_{,zz} + \frac{f 
f_{,R}}{R} + 2 f_{,R}^2 - f_{,z}^2 \right), \label{TRR} \\
T^z_z &=& \frac{c^4}{4 \pi G (1+f)^5 (1-f)} \left(f f_{,RR} + \frac{f 
f_{,R}}{R} + 2 f_{,z}^2 - f^2_{,R} \right), \label{TZZ} \\
T^\varphi_\varphi &=& \frac{c^4}{4 \pi G (1+f)^5 (1-f)} \left[ f 
(f_{,RR} + f_{,zz}) - f^2_{,R} - f^2_{,z} \right], \label{TPHIPHI} \\
T^R_z &=& T^z_R = - \frac{c^4}{4 \pi G (1+f)^5 (1-f)} (f f_{,Rz} - 3 
f_{,R} f_{,z}). \label{TRZ}
\end{eqnarray} 

\noindent In the Newtonian limit, i.e. when $f \ll 1$, equation 
(\ref{TTT}) reduces to the Poisson equation if the function $f$ is 
related to the Newtonian gravitational potential $\Phi$ by $g_{00}=1 + 
2\Phi/c^2$, which means that
\begin{equation}
f=-\Phi/2c^2. \label{fphi}
\end{equation}

\noindent Note that it is a desirable property for every metric to 
obtain a physical consistent newtonian limit, and always we interpret 
the four dimensional metric based on this limit. As an example, we 
interpret the exterior Schwarzschild solution as the space-time solution 
of a spherical symmetric mass body because in its newtonian limit we 
obtain the gravitational potential of a point mass. If the solution does 
not have a consistent newtonian limit then it is, in some sense, 
useless. If we use a potential of the form $\Phi = A/\sqrt{R^2+z^2}$ in 
(\ref{fphi}), where $A$ is a constant, then we obtain the vacuum 
solution.

To find the energy density $\rho$ and the stresses (pressures or 
tensions) of the fluid we must solve the eigenvalue problem for the 
energy-momentum tensor $T^\alpha_\beta V^\beta = \lambda V^\alpha$. In 
this way, we can write the energy-momentum tensor in diagonal form
\begin{eqnarray}
T^{\alpha\beta} = \rho e_{(0)}^\alpha e_{(0)}^\beta + p_{+} 
e_{(1)}^\alpha e_{(1)}^\beta + p_{-} e_{(2)}^\alpha e_{(2)}^\beta + 
p_{\varphi} e_{(3)}^\alpha e_{(3)}^\beta,
\end{eqnarray}

\noindent where
\begin{eqnarray}
\rho = \frac{T^t_t}{c^2}, \hspace{0.5cm}
p_{\pm} = - \frac{T^R_R + T^z_z}{2} \mp 
\frac{1}{2} \sqrt{(T^R_R-T^z_z)^2+4 (T^R_z)^2}, \hspace{0.5cm}
p_\varphi = -T^\varphi_\varphi,
\end{eqnarray}

\noindent and
\begin{eqnarray}
e^\alpha_{(0)} &=& \left(\frac{1+f}{1-f},0,0,0\right), \hspace{0.5cm}
e^\alpha_{(1)} = \left(0,e^R_{(1)},e^z_{(1)},0\right), \\
e^\alpha_{(2)} &=& \left(0,e^R_{(2)},e^z_{(2)},0\right), 
\hspace{0.5cm} e^\alpha_{(3)} = \left(0,0,0,\frac{1}{R (1+f)^2}\right),
\end{eqnarray}

\noindent where
\begin{eqnarray}
e^R_{(1)} &=& -\frac{T^R_z}{(1+f)^2 \sqrt{(T^R_z)^2+(T^R_R+p_{+})^2}}, 
\\
e^z_{(1)} &=& \frac{T^R_R+p_{+}}{(1+f)^2 
\sqrt{(T^R_z)^2+(T^R_R+p_{+})^2}}, \\
e^R_{(2)} &=& -\frac{T^R_z}{(1+f)^2 \sqrt{(T^R_z)^2+(T^R_R+p_{-})^2}}, 
\\
e^z_{(2)} &=& \frac{T^R_R+p_{-}}{(1+f)^2  
\sqrt{(T^R_z)^2+(T^R_R+p_{-})^2}}.
\end{eqnarray}

\noindent In this case the Newtonian effective density is written as 
$\epsilon = \rho+ p_{-}/c^2 + p_{+}/c^2 + p_\varphi/c^2 = \rho - 
T^R_R/c^2 - T^z_z/c^2 - T^\varphi_\varphi/c^2$.  The strong energy 
condition requires that $\epsilon \ge 0$, whereas the weak energy 
condition imposes the condition $\rho \ge 0$. The dominant energy 
condition requires that $|p_R/\rho| \le c^2$, $|p_z/\rho| \le c^2$ and 
$|p_\varphi/\rho| \le c^2$.

Assuming circular geodesic motion of a particle in the equatorial plane 
of the model, we can obtain two interesting physical quantities, i.e. 
the tangential velocity $v_c$, often known as the rotation profile, and 
the angular momentum $h$. This quantities can be written 
as\cite{vog:let3}
\begin{eqnarray}
v_c^2 &=& - \frac{2 c^2 R f_{,R}}{(1-f)(1+f+2Rf_{,R})}, \label{vc} \\ 
h &=& cR^2  (1+f)^2 \sqrt{\frac{-2 f_{,R}}{R[1-f^2+2R f_{,R} (2-f)]}}, 
\label{h}
\end{eqnarray}

\noindent and must be evaluated in $z=0$. The angular momentum written 
above, can be used as an stability criterion for circular orbits on the 
galactic plane by using an extension of the Rayleigh criterion of 
stability of a fluid at rest in a gravitational field\cite{ray,lan:lif}
\begin{eqnarray}
\left. h \frac{dh}{dR} \right|_{z=0} > 0. \label{criterion}
\end{eqnarray}

\section{Ring models}

In a recent work, Vogt \& Letelier\cite{vog:let} presented families of 
flat rings that were constructed by superposing members of the 
Kuzmin-Toomre family of discs. In particular they found a family for 
single flat rings, double flat rings and discs with flat rings. One way 
to generate three-dimensional potential-density from flat ring models is 
to inflate the rings performing the transformation $|z| \rightarrow 
\sqrt{z^2+b^2}$. This is the same transformation used by Miyamoto \& 
Nagai\cite{miy:nag} to inflate the Kuzmin-Toomre discs. In the next 
subsections we construct general relativistic thick models of rings 
based on these flat rings families.

\subsection{Single ring system}

\begin{figure}
\begin{centering}
\epsfig{width=4.cm, height=3.5cm, file=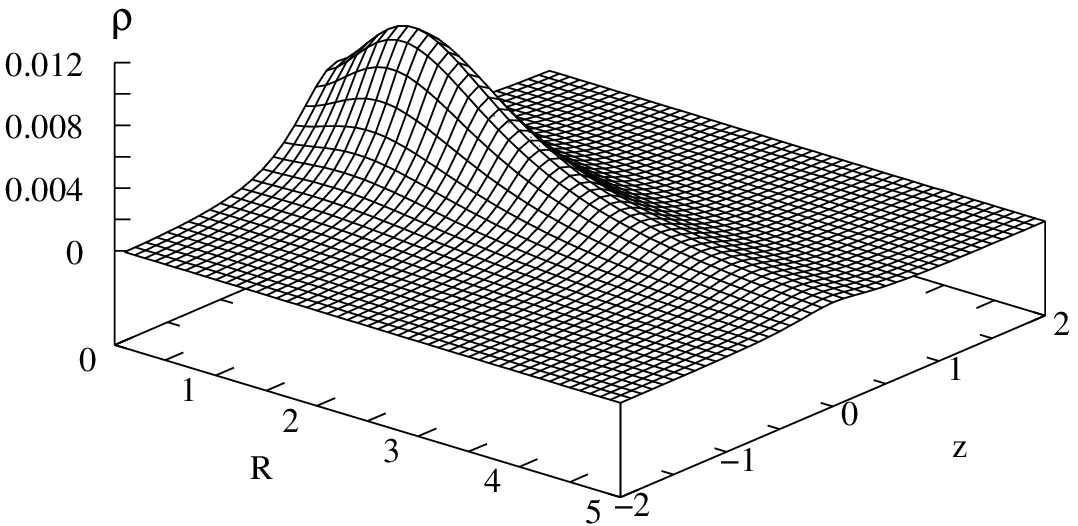}
\epsfig{width=3.5cm, height=3.0cm, file=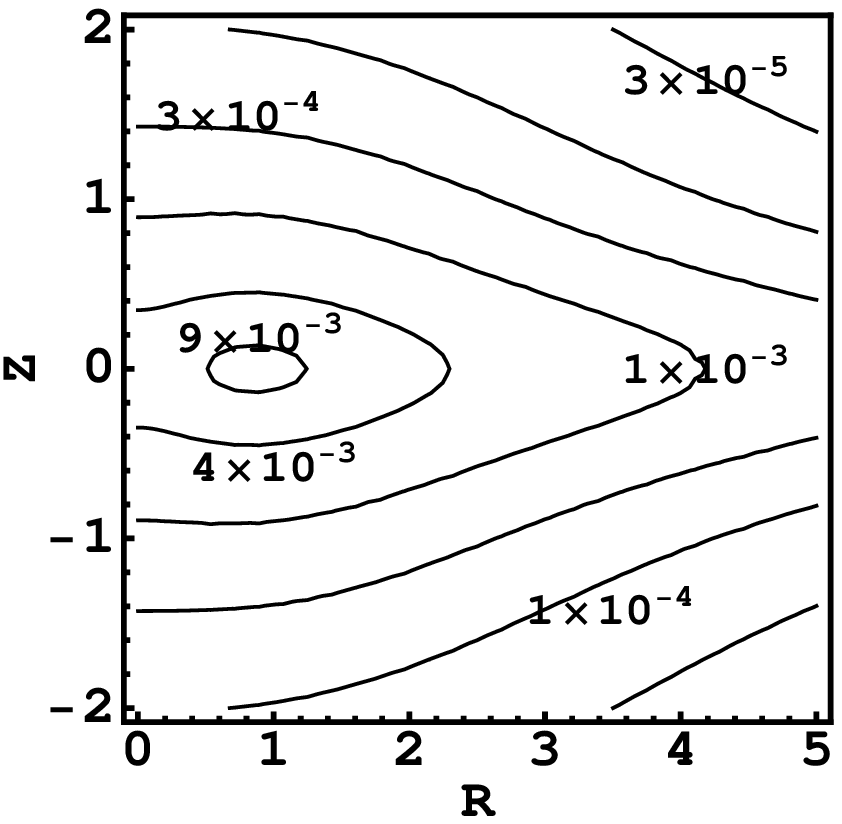} \\
\epsfig{width=4.cm, height=3.5cm, file=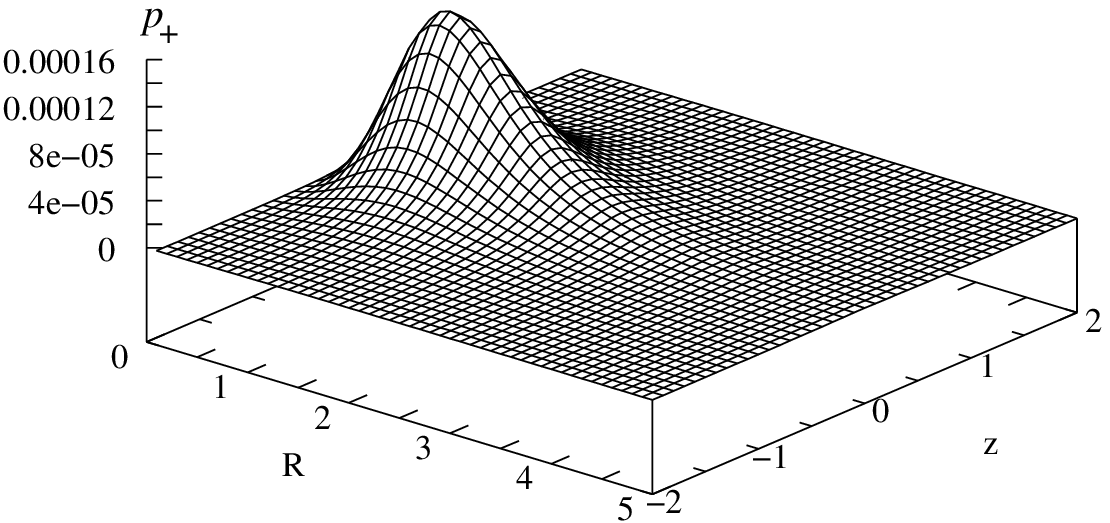}
\epsfig{width=3.5cm, height=3.0cm, file=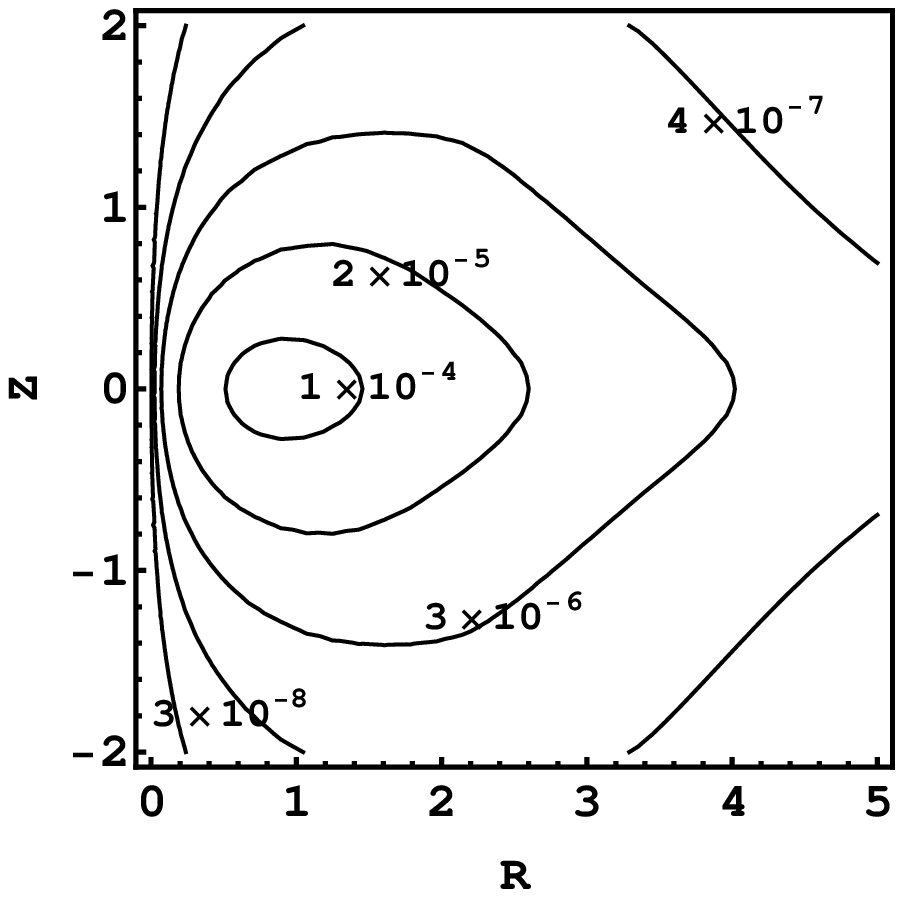} \\
\epsfig{width=4.cm, height=3.5cm, file=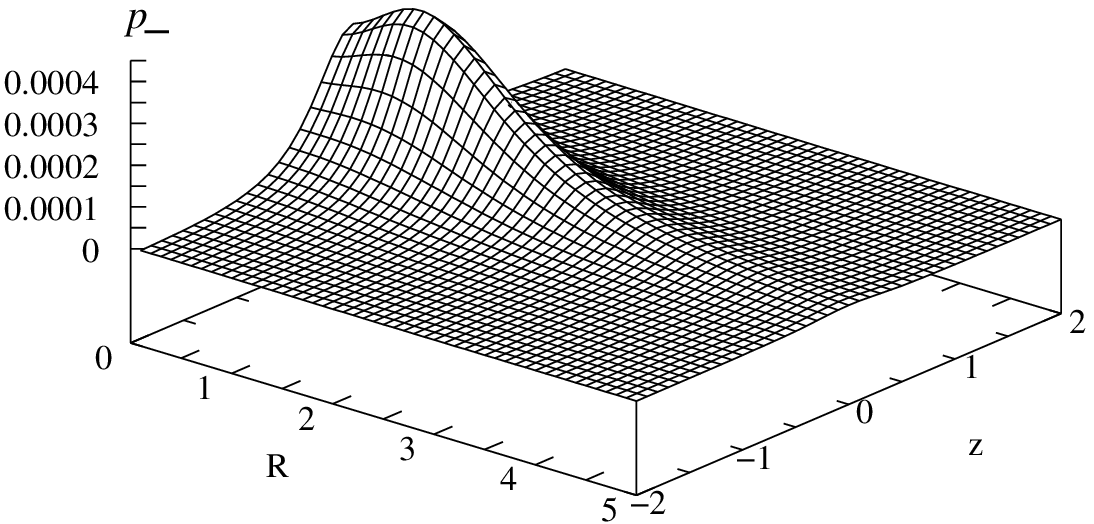}
\epsfig{width=3.5cm, height=3.0cm, file=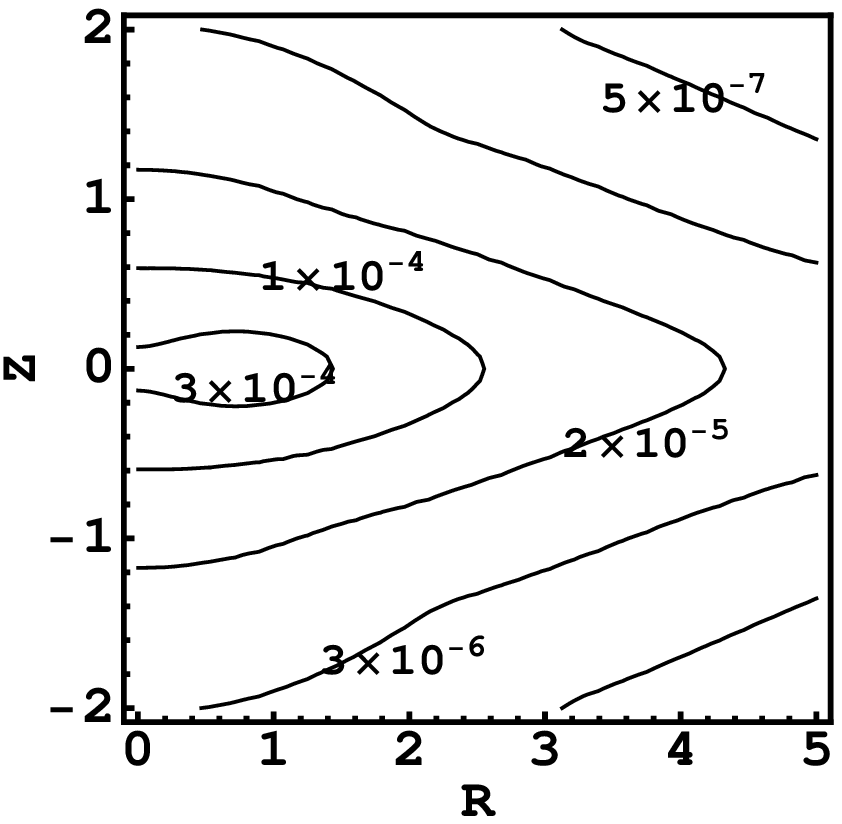} \\
\epsfig{width=4.cm, height=3.5cm, file=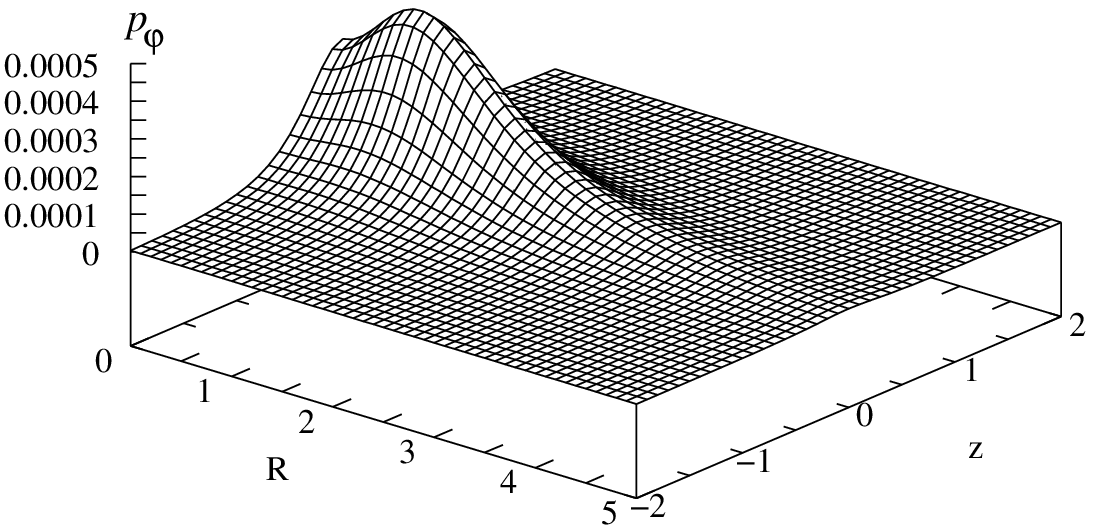}
\epsfig{width=3.5cm, height=3.0cm, file=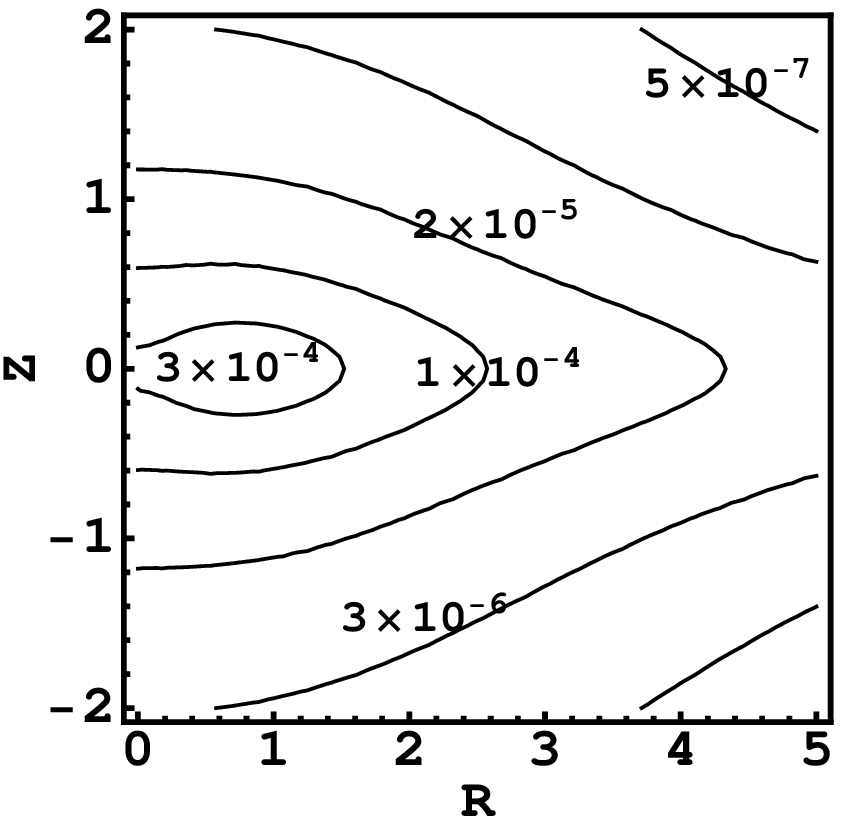} 
\caption{Surface plots and level curves for the energy density and the 
stresses for the case of a general relativistic version of the single 
ring. The parameter values are $a=1$, $b=0.5$ and $\sigma_c=0.1$.}
\label{figure1}
\end{centering}
\end{figure}

\begin{figure}
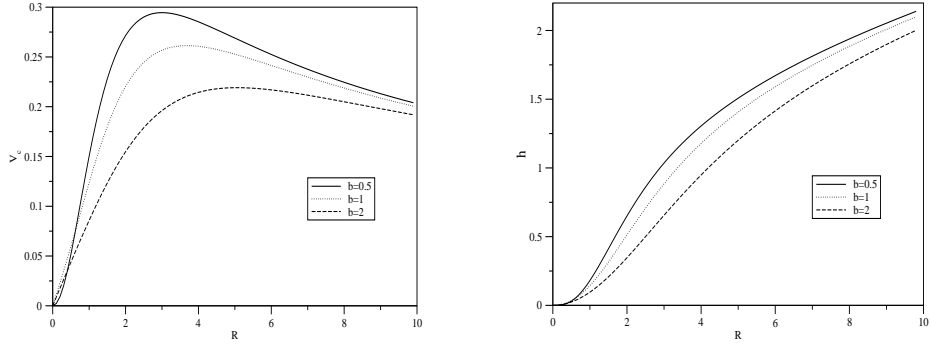

\begin{centering}
\vspace{0.5cm}
\epsfig{width=5.5cm, height=4.5cm, file=figure2a.eps} \hspace{1cm}
\epsfig{width=5.5cm, height=4.5cm, file=figure2b.eps} 
\caption{Tangential velocity and angular momentum for a particle 
orbiting the $z=0$ plane for the same parameter values of Fig. 
\ref{figure1} and $b=0.5,1,2$. The criterion of stability indicates 
that all the orbits depicted are stable.}
\label{figure2}
\end{centering}
\end{figure}

For a general relativistic ring system, we start by taking 
the first member of the family of a single flat ring, which is
\begin{eqnarray}
\phi= -\frac{2 \pi \sigma_c G a^2}{3 [R^2 + (a+ |z|)^2]^{3/2}} 
(2R^2 + 2|z|^2 + a^2 +3a|z|),
\end{eqnarray}

\noindent where $a$ is a constant related to the Kuzmin-Toomre 
potential, $G$ is Newton's gravitational constant, $\sigma_c$ is a 
constant with dimension of surface density and $(R,z)$ are the usual 
cylindrical coordinates. In Ref. \refcite{vog:let} notation, this 
potential corresponds to $\phi^{(1,0)}$, which is the first member of 
the flat single ring potential of Appendix B in the cited article. To 
generate the three dimensional ring model we perform the transformation 
$|z| \rightarrow \sqrt{z^2+b^2}$ and we obtain
\begin{eqnarray}
\Phi = - \frac{2\pi \sigma_c G a^2}{3 \chi^{3/2}}(2 R^2 + 2 \xi^2 
+ a^2 + 3a\xi).
\end{eqnarray}

\noindent where $\xi=\sqrt{z^2+b^2}$ and $\chi=R^2 + (a+\xi)^2$. Note 
that in the above equation we change the notation from $\phi$ to $\Phi$. 
Now, from equation (\ref{fphi}) we obtain the function $f$ and then we 
can find the energy density, $\rho$, and the stresses (pressures or 
tensions), $p_{-}$, $p_{+}$, $p_\varphi$, using the relations 
(\ref{TTT})-(\ref{TRZ}), say
\begin{eqnarray} 
T^t_t &=& \frac{a^2 b^2 c^2 \sigma_c}{2 \xi^3 \chi^{7/2} (1+f)^5} 
\left[ 2 \xi^3 (R^2 + \xi^2) + a^3 (R^2 + 2 \xi^2) + a^2 \xi (7 R^2 + 6 
\xi^2) + a (R^4 \right. \nonumber \\
&& \left. + 8\xi^2 R^2 + 6\xi^4) \right],
 \\ 
T^R_R &=& -\frac{a^4 G \pi \sigma_c^2}{36 \xi^3 \chi^5 (1+f)^5 (1-f)} 
\left\{a^6 \xi^3 + 8 b^2 \xi^3 (R^2 + \xi^2)^2 + 2 a^4 \xi [19 b^4 - 5 
R^2 z^2 \right. \nonumber \\
&& + 3 z^4 + 2 b^2 (5 R^2 +11 z^2)] + a^5 [10 b^4 + 4 z^4 + b^2 (3 R^2 + 
14 z^2)] + 2 a b^2 [19 b^6 \nonumber \\
&& + 3 R^6 + 21 R^4 z^2 + 37 R^2 z^4 + 19 z^6 + b^4 (37 R^2 + 57 z^2) + 
b^2 (21 R^4 + 74 R^2 z^2 \nonumber \\
&&+ 57 z^4)] + a^2 \xi [73 b^6 + 3 b^4 (38 R^2 + 49 z^2) + z^2 (R^4 - 
10 R^2 z^2 + z^4) + b^2 (43 R^4 \nonumber \\
&& + 104 R^2 z^2 + 75 z^4)] + a^3 [72 b^6 + 4 z^4 (-5 R^2 + z^2) + b^4 
(73 R^2 + 148 z^2) \nonumber \\
&& \left. + b^2 (9 R^4 + 53 R^2 z^2 + 80 z^4)]\right\},
 \\
T_z^z &=&-\frac{\pi a^4 G \sigma_c^2}{18 \xi^2 \chi^5 (1+f)^5 (1-f)} 
\left\{-a^6 \xi^2-4 a^5 \xi^3 -2 a^4 \xi^2 (b^2-2 R^2+3 z^2) 
\right. \nonumber \\
&& +4 a^3 \xi [3 b^4 + b^2 (5 R^2+2 z^2) +2 R^2 z^2-z^4] +a^2 [23 b^6+9 
b^4 (4 R^2+5 z^2) \nonumber \\
&& + b^2 (8 R^4 +40 R^2 z^2+21 z^4)-z^2 (R^4-4 R^2 z^2+z^4)] + 4 a b^2 
\xi [4 b^4+b^2 (7 R^2 \nonumber \\
&& \left. + 8 z^2) +3 R^4+7 R^2 z^2+4 z^4] +4 b^2 \xi^2 (R^2+\xi^2)^2 
\right\},
 \\
T_\varphi^\varphi&=&-\frac{a^4 G \pi \sigma_c^2}{36 \xi^3 \chi^5 
(1+f)^5 (1-f)} \left\{a^6 \xi^3 + 8 b^2 \xi^3 (R^2 + \xi^2)^2 + 2 a^4 
\xi [19 b^4 + z^2 (R^2 \right. \nonumber \\
&&+ 3 z^2) + 2 b^2 (8 R^2 + 11 z^2)] + a^5 [10 b^4 + 4 z^4 + b^2 (3 R^2 
+ 14 z^2)] + 2 a b^2 [19 b^6 \nonumber \\
&& + 3 R^6 + 21 R^4 z^2 + 37 R^2 z^4 + 19 z^6 + b^4 (37 R^2 + 57 z^2) + 
b^2 (21 R^4 + 74 R^2 z^2 \nonumber \\
&& + 57 z^4)] + a^2 \xi [73 b^6 + z^2 (R^2 + z^2)^2 + 21 b^4 (6 R^2 + 7 
z^2) + b^2 (43 R^4 + 128 R^2 z^2 \nonumber \\
&& + 75 z^4)] + a^3 [72 b^6 + 4 z^4 (R^2 + z^2) + b^4 (97 R^2 + 148 
z^2) + b^2 (9 R^4 + 101 R^2 z^2 \nonumber \\
&& \left. + 80 z^4)]\right\}, 
 \\
T_z^R&=&T^z_R= \frac{a^6 G \pi R z \sigma_c^2}{6 \xi \chi^5 (1+f)^5 
(1-f)} [a^3 -R^2 (a+\xi) + \xi (3 a^2 + \xi^2 + 3 a \xi)].
\end{eqnarray}

In this case the tangential velocity and the angular momentum for a 
particle in circular orbits are
\begin{eqnarray} 
v_c^2 &=& \frac{2 \pi G \sigma_c a^2 c^2 R^2 (2R^2 + 2\xi^2 +a\xi 
-a^2)}{(1-f) [3 c^2 \chi^{5/2} (1+f) - 2 \pi G \sigma_c a^2 R^2 
(2R^2+2\xi^2 + a\xi -a^2)]}, \\
h &=& c R^2 (1+f)^2 \nonumber \\
&& \times \sqrt{\frac{2 \pi G a^2 \sigma_c (2R^2 + 2\xi^2 + 
a\xi -a^2) }{3 c^2 \chi^{5/2} (1-f^2) - 2 \pi G \sigma_c a^2 R^2 
(2R^2 + 2\xi^2 +a\xi -a^2) (2-f)}}.
\end{eqnarray}

\noindent where the values of $\xi$ and $\chi$ have to be evaluated in 
the plane $z=0$.

In Fig. \ref{figure1} we present the surface and level curves of the 
energy density and the stresses for the parameters $a=1$, $b=0.5$ and 
$\sigma_c=0.1$. Hereafter we used $G=c=1$ for the plots and numerical 
calculations. As in the Newtonian case, the lower the ratio $b/a$ the 
flatter the mass distribution. Numerically we verified, that for these 
values $\rho>0$ and $\epsilon>0$, moreover $|p_{+}/\rho| < 0.017$, 
$|p_{-}/\rho| < 0.051$ and $|p_{\varphi}/\rho| < 0.051$, which means 
that the strong, weak and dominant energy conditions are satisfied. From 
the energy density profile and level curves we see that we have a ring 
structure. In Fig. \ref{figure2} we show the tangential velocity and the 
angular momentum for a particle in circular orbit for the same values of 
Fig. \ref{figure1} and different values of the parameter $b$. Applying 
the criterion of stability (\ref{criterion}) we see that these orbits 
are stable. For lower values of the parameter $b$, e.g. $b=0.4$, the 
stresses become negative, first near $R\approx 0$ and $z\approx 0$ and 
latter spreading to nearest values of $R$ and $z$. This negatives values 
characterized tensions instead of pressures. The stability criterion 
indicates that, in this cases, the orbits in the plane $z=0$ are 
non-stable in the region where we have tensions.

\subsection{Disc with a ring}

\begin{figure} 
\begin{centering} 
\epsfig{width=4.cm, height=3.5cm, file=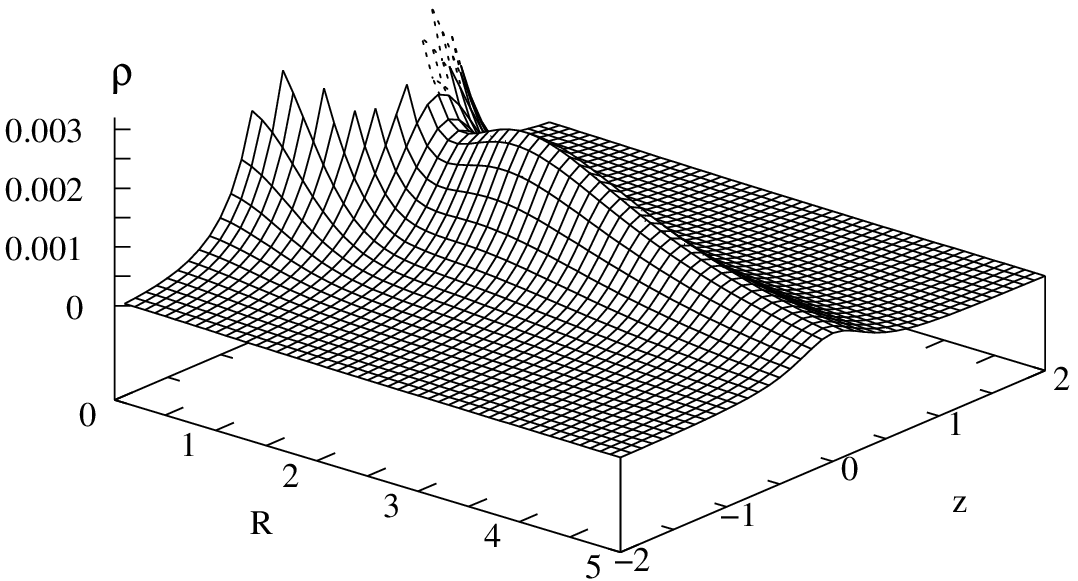} 
\epsfig{width=3.5cm, height=3.0cm, file=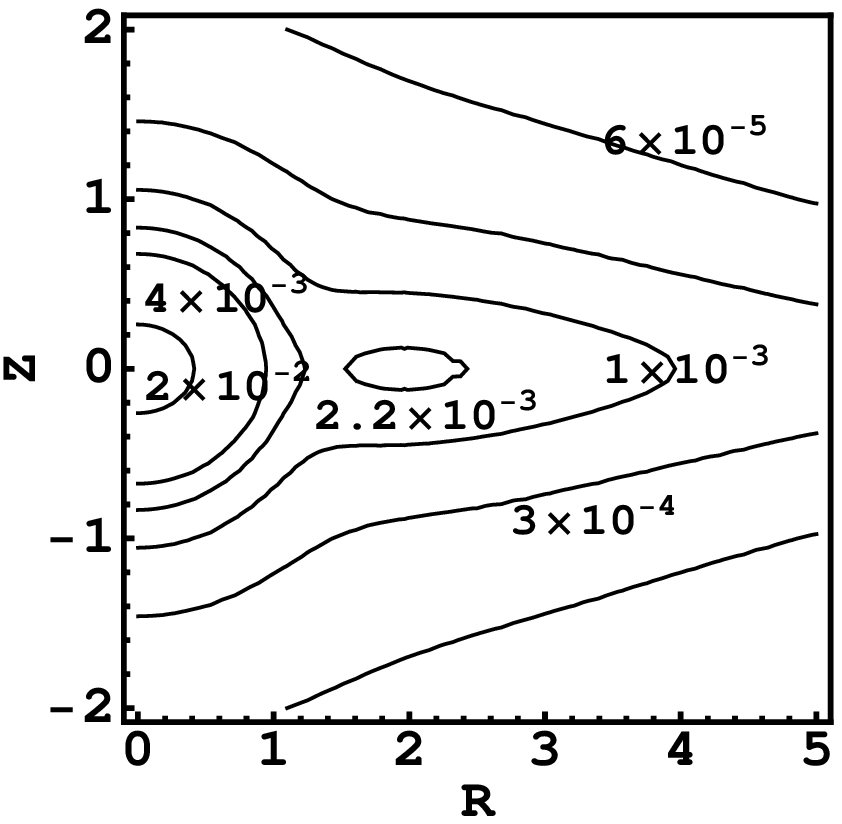} \\
\epsfig{width=4.cm, height=3.5cm, file=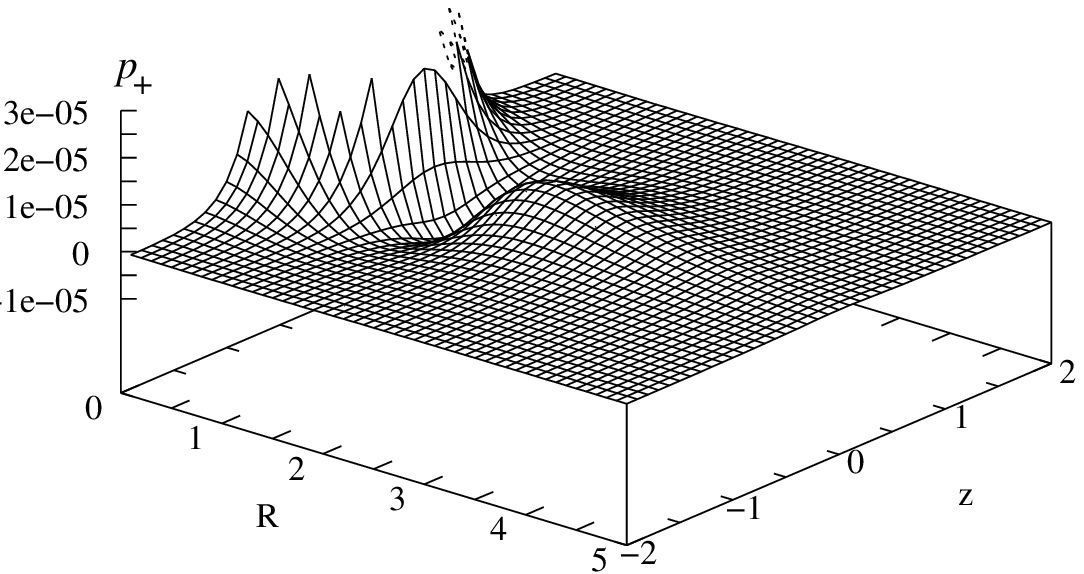} 
\epsfig{width=3.5cm, height=3.0cm, file=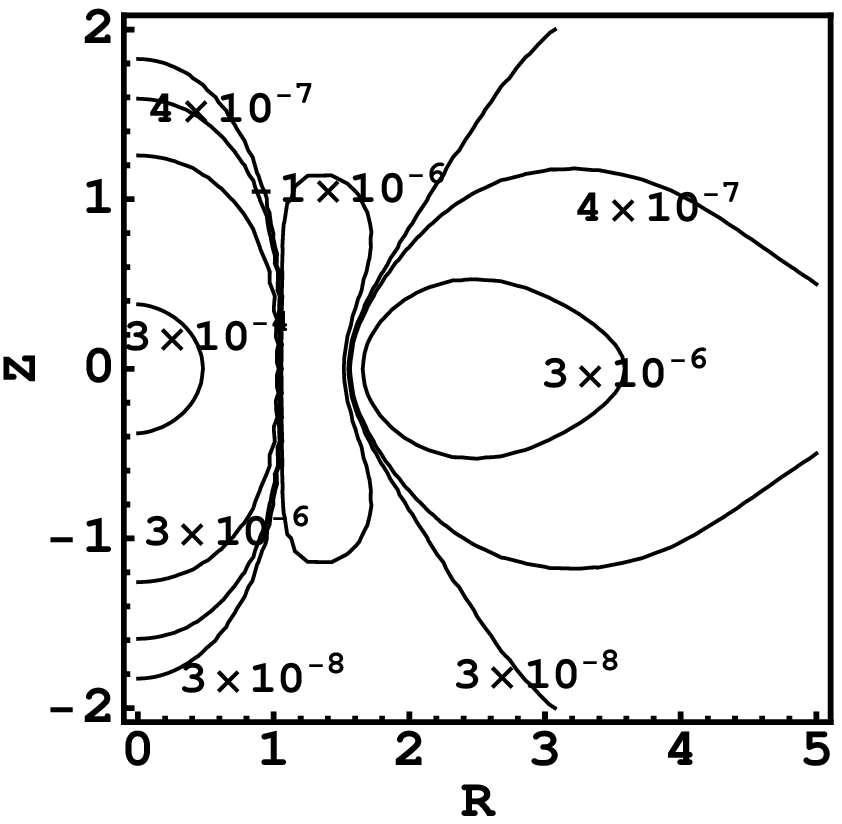} \\ 
\epsfig{width=4.cm, height=3.5cm, file=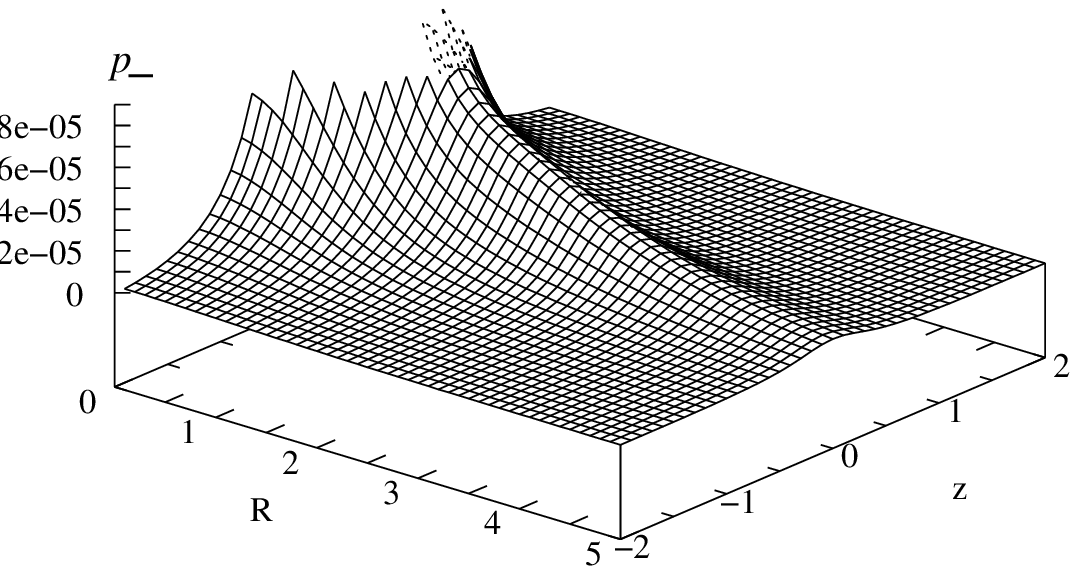} 
\epsfig{width=3.5cm, height=3.0cm, file=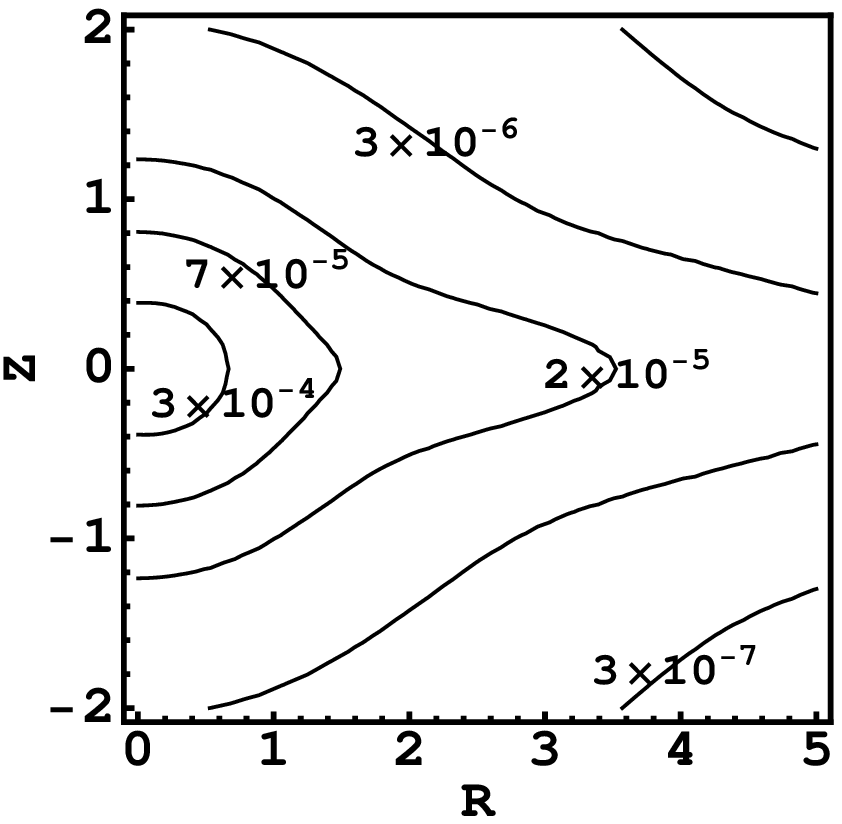} \\
\epsfig{width=4.cm, height=3.5cm, file=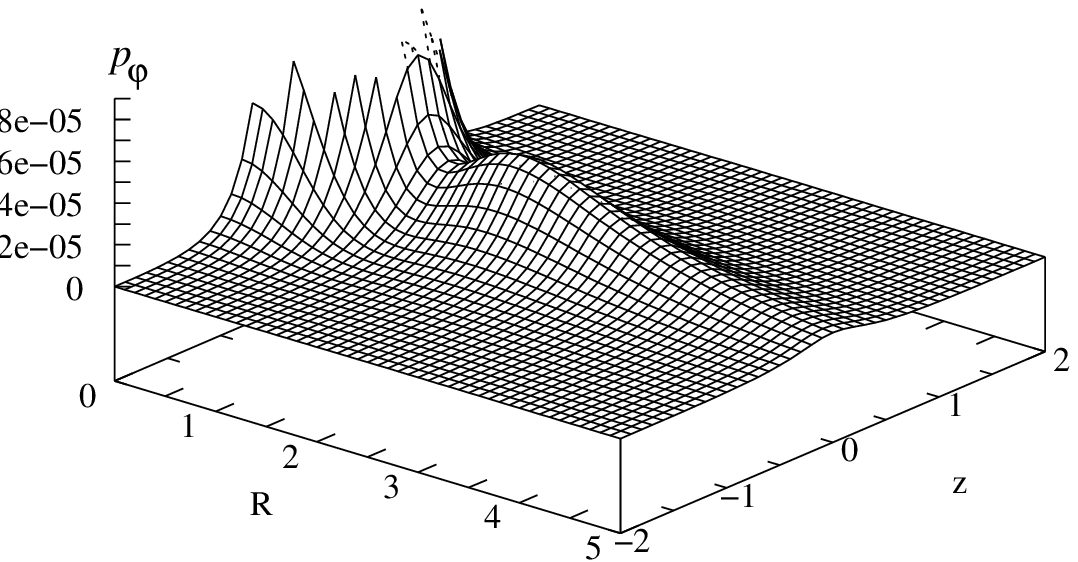} 
\epsfig{width=3.5cm, height=3.0cm, file=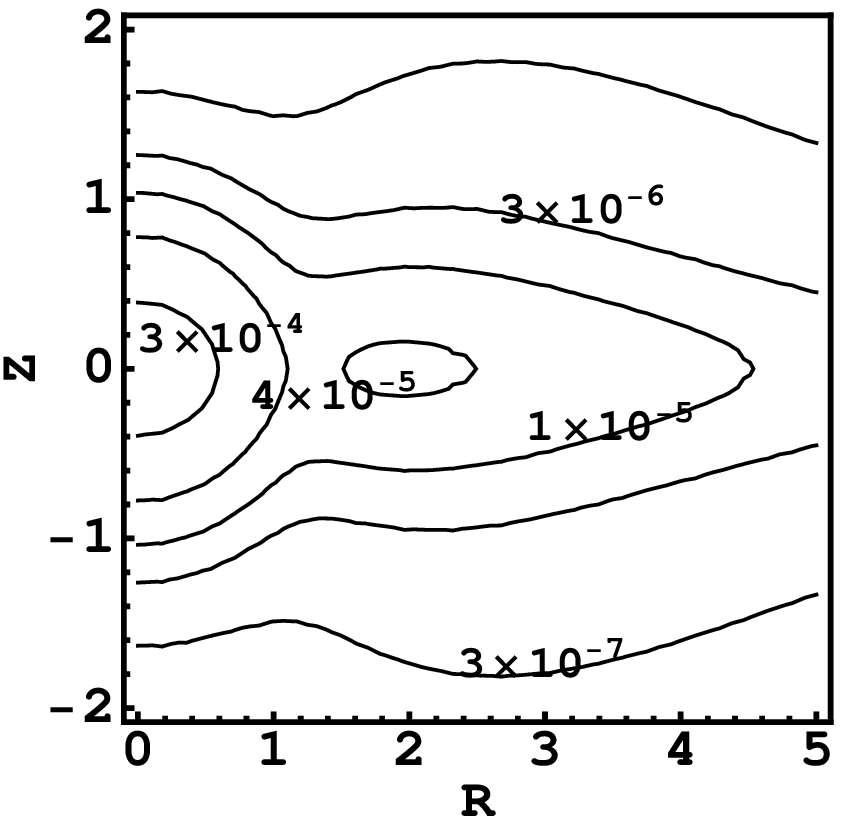} 
\caption{Surface plots and level curves for the energy density and the 
stresses for the general relativistic case of a disc with a ring. The 
parameter values are $a=k=1$, $b=0.5$ and $\sigma=0.1$. The graph are 
plot in such a way to enhance the ring peak and not the central 
disc peak. Note that the stress $p_{+}$ has negative values in the 
intersection of the disc and the ring.} 
\label{figure3}
\end{centering} 
\end{figure}

\begin{figure}
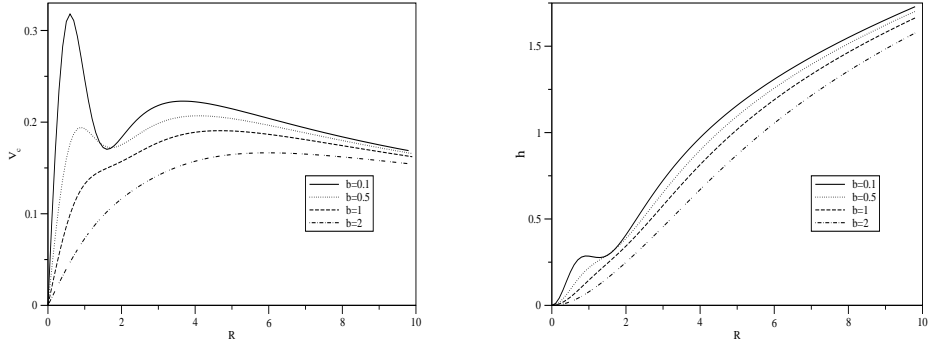

\vspace{0.5cm}
\begin{centering}
\epsfig{width=5.5cm, height=4.5cm, file=figure4a.eps} \hspace{1cm}
\epsfig{width=5.5cm, height=4.5cm, file=figure4b.eps} 
\caption{Tangential velocity and angular momentum for a particle 
orbiting the $z=0$ plane for the same parameter values of Fig. 
\ref{figure3} and $b$=0.1, 0.5, 1, 2. The angular momentum curve with 
$b=0.1$ has a region of non-stability near the radius 1.5 that 
corresponds to the valley at the same radius in the tangential velocity 
graph.}
\label{figure4}
\end{centering}
\end{figure}

Following the previous section, we start with the Newtonian potential 
that represent a disc with a flat ring, say
\begin{eqnarray}
\phi &=&  -\frac{2 \pi \sigma_c G a^2}{15 [R^2+ (a+|z|)^2]^{5/2}} 
\left\{ [R^2+ (a+|z|)^2]^2(3-4 k^2)+ [R^2+ (a+|z|)^2] \right. \nonumber 
\\
&& \times [a (a +|z|)(3-4 k^2) + k^4 (8R^2 + 8|z|^2 + 9 a |z|)] + a^2 
[-R^2 (1+2k^2) + (a+|z|)^2 \nonumber \\
&& \left. \times (2 + 4k^2 + 3k^4)] \right\},
\end{eqnarray}

\noindent where $k$ is a constant related to the disk radius. This 
potential corresponds to the $\phi^{(2)}_{(d)}$ potential of 
\cite{vog:let}. To generate the three dimensional ring model we perform 
the transformation $|z| \rightarrow \sqrt{z^2+b^2}$ and we obtain
\begin{eqnarray}
\Phi &=& -\frac{2 \pi \sigma_c G a^2}{15 \chi^{5/2}} \left\{ \chi^2(3-4 
k^2)+ \chi[a (a+\xi)(3-4 k^2) + k^4 (8R^2 + 8\xi^2 + 9 a \xi)] \right. 
\nonumber \\
&& \left. + a^2 [-R^2 (1+2k^2) + (a+\xi)^2(2 + 4k^2 + 3k^4)] \right\} 
\label{discring}
\end{eqnarray}

\noindent where $\xi$ and $\chi$ are defined as in the previous section. 
The energy density and the stresses are obtained from equations 
(\ref{TTT})-(\ref{TRZ}) and relation (\ref{fphi}). Then, the tangential 
velocity and the angular momentum of a circular orbital in the plane of 
the discs are found from (\ref{vc}) and (\ref{h}). The analytical 
expressions for all these quantities are cumbersome and will be omitted. 
In Fig. \ref{figure3} we plot the surface and level curves of the energy 
density and the stresses for the model of equation (\ref{discring}) with 
parameter values $a=k=1$, $b=0.5$ and $\sigma_c=0.1$. Numerically we 
found that $\rho>0$, $\epsilon>0$, $|p_{+}/\rho| < 0.024$, 
$|p_{-}/\rho|< 0.043$ and $|p_\varphi/\rho| < 0.028$, and therefore all 
the energy conditions are met. In the energy density graph, the figure 
is plot in such a way to enhance the region in which we clearly see the 
ring peak and not the disc central peak. For lower values of the 
parameter $b$ the disc and the ring became narrower. Note that in the 
$p_{+}$ graph, the stress is negative in the intersection between the 
disc and the ring representing tensions and not pressures. The other 
stresses, $p_{-}$ and $p_{\varphi}$, are always positive. In Fig. 
\ref{figure4}, we plot the tangential velocity and the angular momentum 
for a circular orbiting particle for the same parameter values of Fig. 
\ref{figure3} and different values of the parameter $b$. Using the 
stability criterion (\ref{criterion}), we see that for $b=0.1$ the 
particle orbit is not stable for radius between $R \approx 1$ and $R 
\approx 1.5$. For higher values of $b$ the circular orbits are stable. 
We check also that for lower values of $b$ (thinner systems) regions of 
non-stability appear. This non-stability corresponds to the valley in 
the tangential velocity graph, and also to the radius where the stress 
$p_{+}$ has negative values. Moreover, we find that for lower values of 
the parameter $b$ the negative value region in the stress $p_{+}$ 
becomes larger.

\subsection{Double ring system}

\begin{figure}
\begin{centering}
\epsfig{width=4.cm, height=3.5cm, file=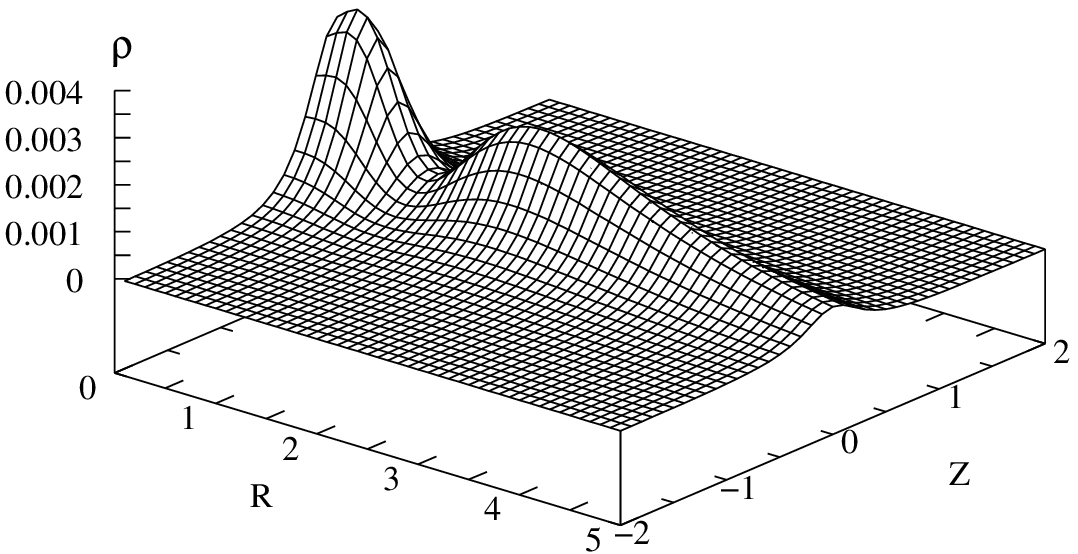}
\epsfig{width=3.5cm, height=3.0cm, file=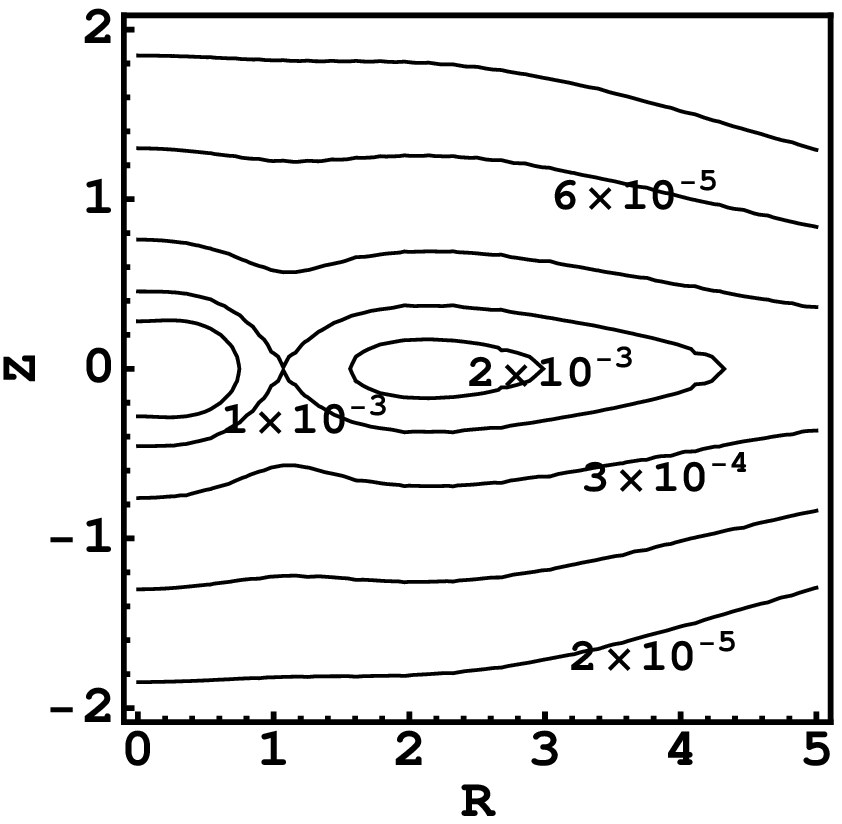} \\
\epsfig{width=4.cm, height=3.5cm, file=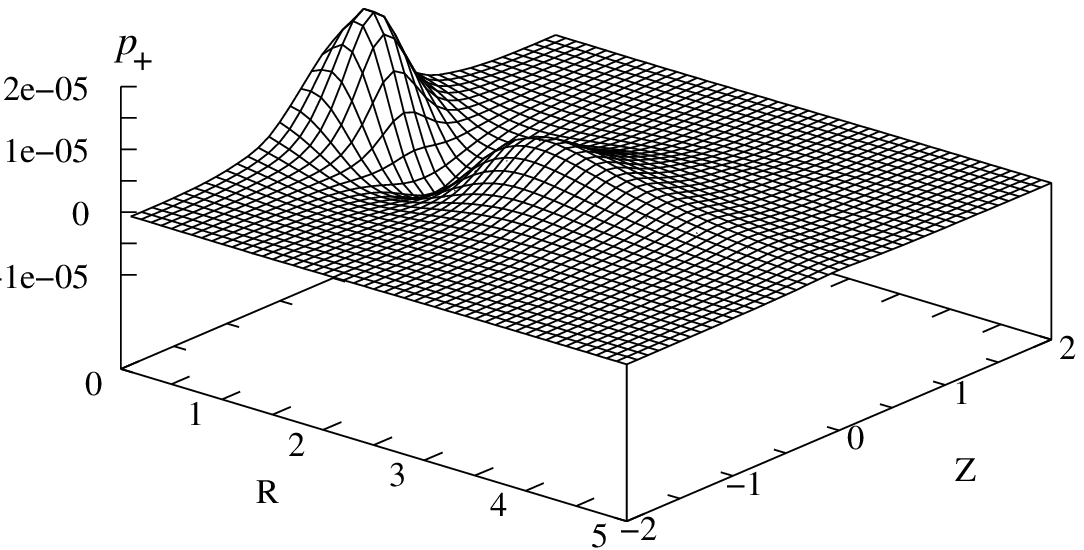}
\epsfig{width=3.5cm, height=3.0cm, file=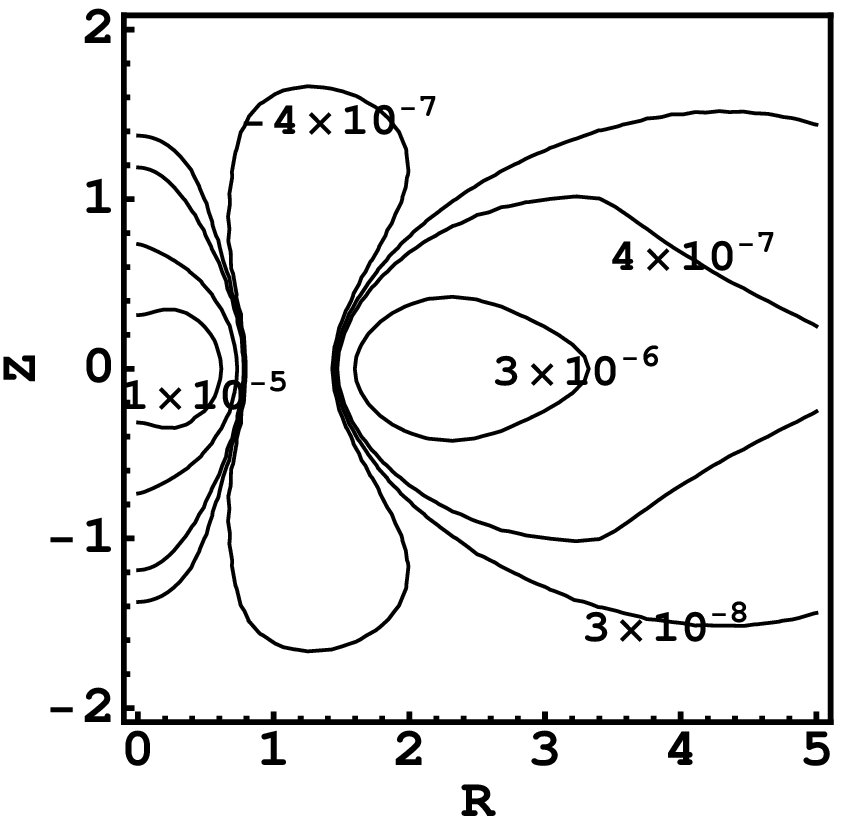} \\
\epsfig{width=4.cm, height=3.5cm, file=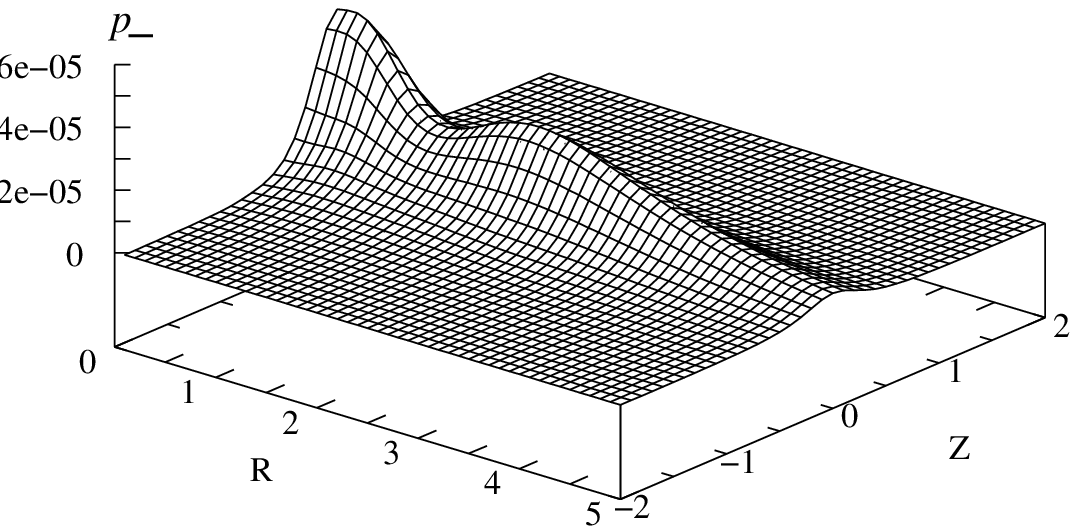}
\epsfig{width=3.5cm, height=3.0cm, file=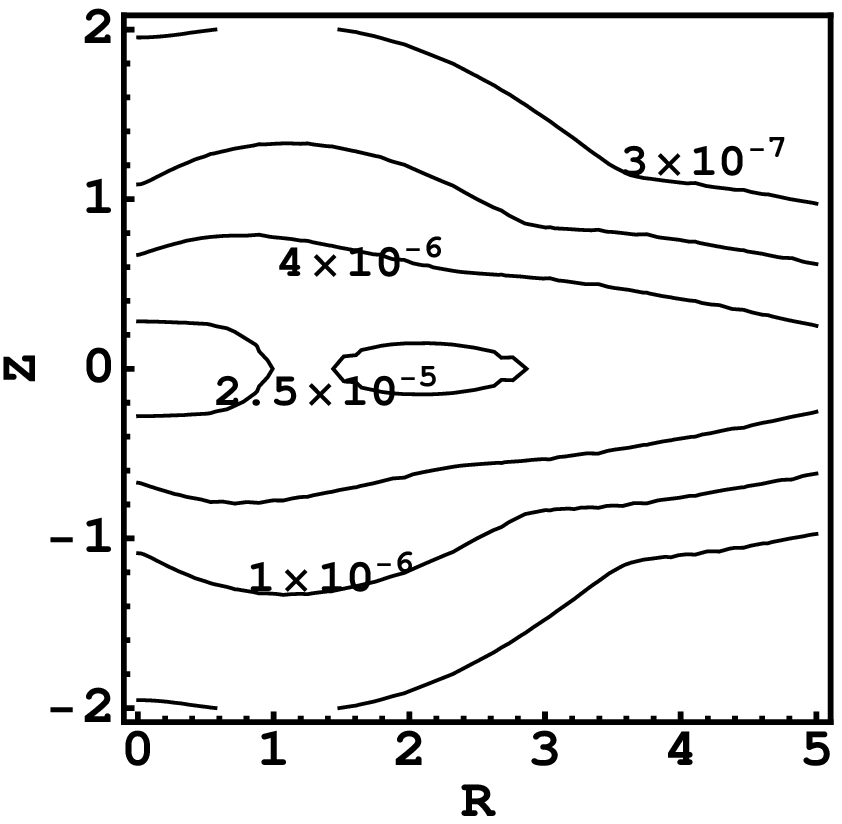} \\
\epsfig{width=4.cm, height=3.5cm, file=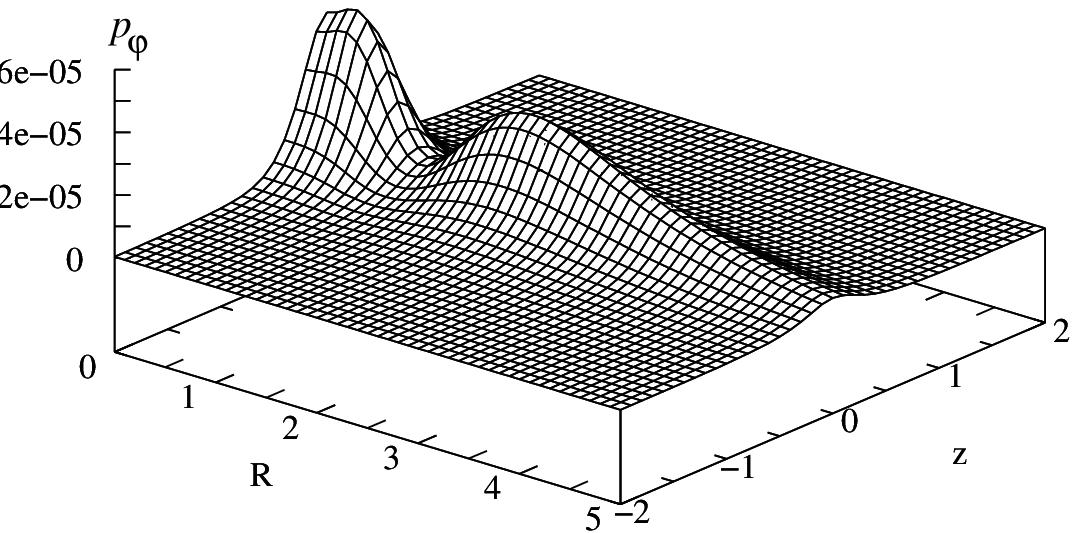}
\epsfig{width=3.5cm, height=3.0cm, file=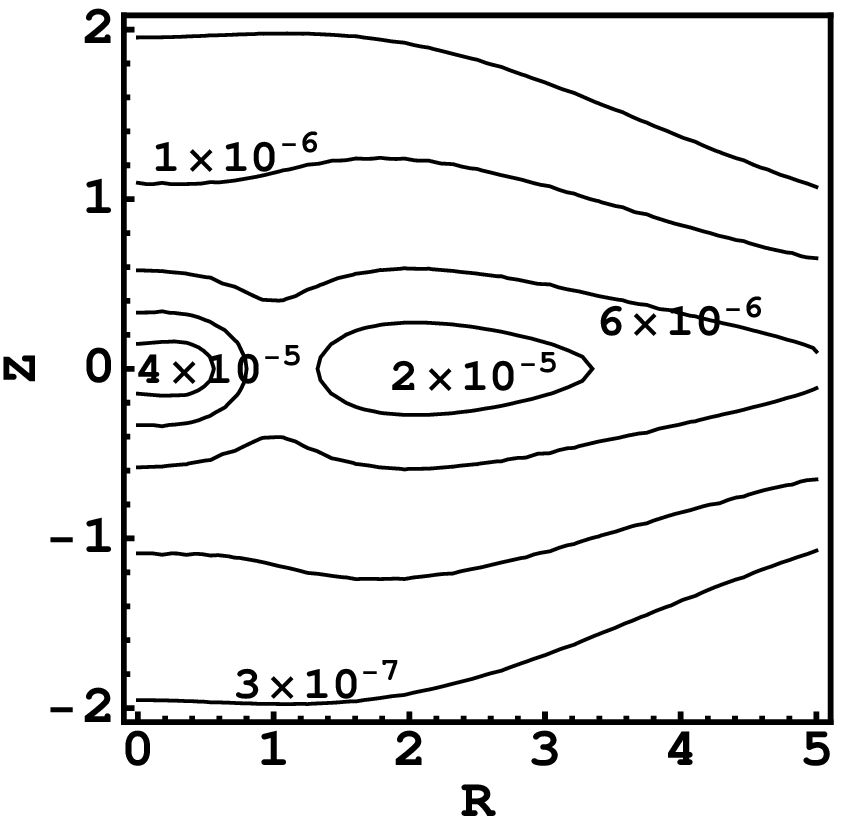} 
\caption{Surface plots and level curves for the energy density and the 
stresses for the general relativistic case of a two ring system. The 
parameter values are $a=k=1$, $b=0.5$ and $\sigma=0.1$. Note 
that the stress $p_{+}$ has negative values in the 
intersection of the rings.} 
\label{figure5}
\end{centering}
\end{figure}

\begin{figure}
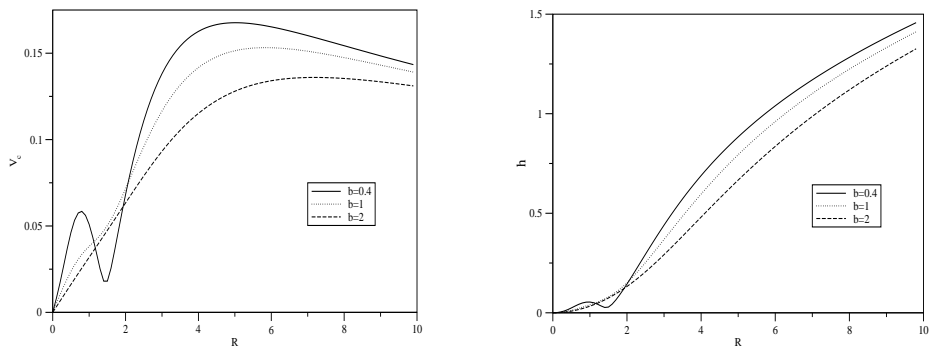

\vspace{0.5cm}
\begin{centering}
\epsfig{width=5.5cm, height=4.5cm, file=figure6a.eps} \hspace{1cm}
\epsfig{width=5.5cm, height=4.5cm, file=figure6b.eps} 
\caption{Tangential velocity and angular momentum for a particle 
orbiting the $z=0$ plane for the same parameter values of Fig. 
\ref{figure5} and $b$=0.4, 1, 2. The angular momentum curve with $b=0.4$ 
has a region of non-stability from $R\approx 1$ to $R\approx1.5$.}
\label{figure6}
\end{centering}
\end{figure}

In this case, the Newtonian potential to be consider is the first member 
of the family of double rings find in Ref. \refcite{vog:let}, which is
\begin{eqnarray}
\phi &=& \frac{-2 \pi \sigma_c G a^2}{105 [R^2+(a+|z|)^2]^{7/2}} 
\left\{6 [R^2+(a+|z|)^2]^3 (1+8k^4) + [R^2+(a+|z|)^2]^2 \right. 
\nonumber \\ 
&& \times [-16 k^2 (R^2+|z|^2) +a^2(9- 26k^2 -54k^4) + 3a |z| 
(2-16k^2-19k^4)] -a^2 [R^2 \nonumber \\
&& + (a+|z|)^2] [2 R^2 (2+ 11k^2 + 9k^4) + |z|^2 (1- 26k^2 - 27k^4) - a 
|z| (13+82k^2 \nonumber \\
&& + 69 k^4) -2 a^2 (7 +28k^2 +21 k^4)] - 3 a^3(a+ |z|) \left. 
(1+k^2)^2 [2R^2 + 7(a+|z|)^2]\right\}, \nonumber \\
\end{eqnarray}

\noindent where $k$ is a constant related to the location of the gap 
between the rings. This potential corresponds to the $\phi^{(2)}_{(d)}$ 
potential of Ref. \refcite{vog:let}. Performing the previously applied 
transformation $|z| \rightarrow \sqrt{z^2+b^2}$ we obtain
\begin{eqnarray}
\Phi &=& \frac{-2 \pi \sigma_c G a^2}{105 \chi^{7/2}} \left\{6 \chi^3 
(1+8k^4) + \chi^2[-16 k^2 (R^2+\xi^2)+a^2(9- 26k^2 -54k^4) 
\right. \nonumber \\ 
&& + 3a \xi (2 -16k^2-19k^4)] - a^2 \chi [2 R^2 (2+ 11k^2 + 9k^4) + 
\xi^2 (1- 26k^2 - 27k^4) \nonumber \\ 
&&  - a \xi (13+82k^2 +69k^4) -2a^2(7 +28k^2 +21 k^4)] - 3 a^3(a+ 
\xi) (1+k^2)^2 \nonumber \\ 
&& \left. \times [2R^2 + 7(a+\xi)^2]\right\},
\end{eqnarray}

\noindent where $\xi$ and $\chi$ are defined as in the previous section. 
The exact expressions for the energy density, the stresses, the 
tangential velocity and the angular momentum, are obtained 
straightforward from equations (\ref{TTT})-(\ref{TRZ}), (\ref{vc}) and 
(\ref{h}) using (\ref{fphi}). In Fig. \ref{figure5}, we plot the surface 
and level curves of the energy density and the stresses for the double 
ring system for parameter values $a=k=1$, $b=0.5$ and $\sigma_c=0.1$. 
For this parameter values we found that $\rho>0$, $\epsilon>0$, 
$|p_{+}/\rho| < 0.028$, $|p_{-}/\rho|< 0.023$ and $|p_\varphi/\rho| < 
0.023$, and the model satisfies all the energy conditions. The energy 
density graph shows two ring peaks of approximately the same magnitude. 
The $p_{+}$ stress has negative values between the rings, this is 
similar to what we find in the disc with a ring case. The other stresses 
are positive. As in the Newtonian case, for lower values of the ratio 
$b/a$ the two rings became thinner. In Fig. \ref{figure6}, we plot the 
tangential velocity and the angular momentum of the particle with the 
same parameter values of Fig. \ref{figure5} and different values of the 
parameter $b$. We see from the angular momentum graph, that for $b=0.4$ 
the circular orbits are non-stable in a region between $R\approx 1$ and 
$R\approx 1.5$. Note that this correspond, as in the disc with a ring 
case, to the valley located at the same radius in the tangential 
velocity profile for the same value of $b$. But, the tangential velocity 
curves are a bit different from the disc and a ring case because the 
starting values of the monotonically decreasing region ($R>4$ 
approximately) are higher than the initial bump. For higher values of 
the parameter $b$ the circular orbits are always stable. Also, we find 
that when we lower the parameter $b$, the negative value region in the 
stresses increases.

\section{Conclusions}

In this work we presented several ring model systems in the context of 
general relativity. This was achieved by inflating previously 
constructed Newtonian ring potentials using the transformation $|z| 
\rightarrow \sqrt{z^2+b^2}$, and then finding their relativistic 
analogs. We use an inverse method because, in general, the direct method 
for solving Einstein's field equations for more sophisticated models is 
difficult and cumbersome, and in most of the cases the solutions do not 
have a clear physical interpretation at the newtonian limit, making 
these solutions, in some sense, useless. In the other hand, if we want 
to solve the Einstein field equations using the direct method for a more 
sophisticated model, say several concentric rings of matter, we do not 
know how to construct the correct form of the energy momentum tensor of 
our system, and maybe after solving it we do not have a clear physical 
interpretation at the newtonian limit.

The models presented in this work have infinite extension but the 
physical quantities decays very fast with the distance, and in 
principle, one could make a cut-off radius to consider it finite. The 
different systems studied present a region of non-stability in the 
intersection of the disc and the ring, and between the rings. This 
region of non-stability is present in thinner systems and corresponds to 
a region in which one of the stresses is negative, i.e. we have tensions 
instead of pressures. A better approach to the stability of the system 
has to take into account the collective behavior of the particles and 
not only a geodesic approximation. This will be done in a future work. 
Some examples of general relativistic stability analysis of discs can be 
found in Ref. \refcite{uje:let,uje:let2,uje:let3}.

In summary, this article shows several general relativistic ring systems 
that are physical acceptable and have a clear interpretation at the 
newtonian limit, and from which we can test general relativistic 
effects.

\section*{Acknowledgements}

M.U. and P.S.L. thanks CNPq for financial support; P.S.L. also thanks 
FAPESP; D.V. thanks FAPESP for financial support.

\end{document}